\documentclass[prd,reprint]{revtex4-1}
\usepackage{amsmath}
\usepackage{graphicx}
\usepackage[usenames]{color} 
\usepackage{ulem}
\expandafter\let\csname equation*\endcsname\relax
\expandafter\let\csname endequation*\endcsname\relax

\begin{document}
\title[CCDs for detection of coherent neutrino-nucleus scattering]{Charge Coupled Devices for detection of coherent neutrino-nucleus scattering}

\author{Guillermo Fernandez Moroni$^{1,2,3}$, Juan Estrada$^3$, Gustavo Cancelo$^3$, \\Eduardo Paolini$^{2,4}$, Javier Tiffenberg$^3$, Jorge Molina$^5$}
\address{$^1$Comisi\'{o}n de Investigaciones Cient\'{i}ficas y T\'{e}cnicas (CONICET), Argentina}
\address{$^2$Universidad Nacional del Sur, Av. Alem 1253, (8000) Bah\'{i}a Blanca, Argentina}
\address{$^3$Fermi National Accelerator Laboratory, Batavia IL, United States.}
\address{$^4$Comisi\'{o}n de Investigaciones Cient\'{i}ficas Prov. Buenos Aires (CIC), Buenos Aires, Argentina}
\address{$^5$Universidad Nacional de Asunci\'{o}n, Asunci\'{o}n, Paraguay}

\begin{abstract}

This article details the potential for using Charge Coupled Devices (CCD) to 
detect low-energy neutrinos through their coherent scattering with nuclei.
The detection of neutrinos through this standard model process has not 
been accessible because of the small energy deposited in such
interactions with the detector nuclei. Typical particle detectors have thresholds 
of a few keV, and most of the energy deposition expected from coherent scattering
is well below this level. The devices we discuss can be operated 
at a threshold of approximately 30 eV, making them ideal for observing this signal. 
For example, the number of coherent scattering events expected on a 52 gram CCD array located 
next to a power nuclear reactor is estimated to be near to 626 events$/$year.  The results of our study show that  
detection at a confidence level of 99$\%$ can be reached within three months for this kind of
detector array.
\end{abstract}

\pacs{1315, 9440T}

\maketitle

\section{INTRODUCTION}

Since the discovery of neutral-current neutrino interactions in 1973 by Hasert \textit{et al.} \cite{Hasert 1973}, the importance of the coherent enhancement in elastic neutrino scattering has been pointed out \cite{Freedman 1974}, along with its implication for studies of star collapse. Unfortunately, it is difficult to detect because of its very small cross section  ($< 10^{-39}$ cm$^2$) \cite{Freedman 1977} and the small energy deposition, typically less than $<$10 keV for any material. Detector technology has not met yet the extreme requirements on detector mass or on the energy threshold. Nevertheless, in recent times, interest from low energy neutrino physics has been increasing, mainly for verifying predictions of the standard model (SM), and exploring the possibilities of new physics beyond the SM at very small energy scales \cite{Scholberg 2006}. In astrophysics, for example, the understanding of MeV-neutrino physics has great relevance for energy transport in supernovas and it is related to the ongoing effort to develop new supernova detectors. These kind of detectors can also be used to monitor nuclear reactors through their emitted neutrinos \cite{Barbeau 2003, Hagmann 2004}.

Although initially devised as memory devices \cite{Boyle 2010, Smith 2010}, CCD have found a niche as imaging detectors due to their ability to obtain high resolution digital images of objects placed in their line of sight. In particular, scientific CCDs have been used extensively in ground and space-based astronomy and X-ray imaging \cite{Janesick 2001}. These devices have high detection efficiency, low noise, good spatial resolution, and low dark current. Furthermore, thick CCDs with increased detection mass enable their use as particle detectors \cite{Holland 2003}. Using this technology, the DAMIC search for cold dark matter has been deployed at Snolab \cite{Barreto 2012}. %

Several nuclear-reactor neutrino experiments were based mostly on inverse beta decay \cite{Double Chooz,Boehm 2000,kamland}, usually using large volumes of target materials to counter their relatively high threshold energy of several keV. Recently, with the decreasing threshold of solid-state detectors there has been a growing interest in using them for neutrino detection \cite{Xim 2005,Wong 2007}. In this paper, we discuss the potential for using CCD technology for neutrinos scattering. We analyze the potential detection of neutrinos at a detector threshold of 28 eV of ionizing energy (five times larger than the RMS noise of 5.5 eV). The low-energy threshold of a CCD provides an opportunity to detect the main mechanism of neutrino-nucleus coherent scattering, which has never been observed. The proposed detector uses a mass of approximately 100 g of Si, which allows the construction of a small-sized nuclear-reactor neutrino detector. Our focus is on neutrinos with energies of $<$ 12 MeV produced at a nuclear reactor. %

\section{HIGH RESISTIVITY SCIENTIFIC CCD}
\label{sec: high resitivity CCDs}

\begin{figure}
\resizebox{1\hsize}{!}{\includegraphics*{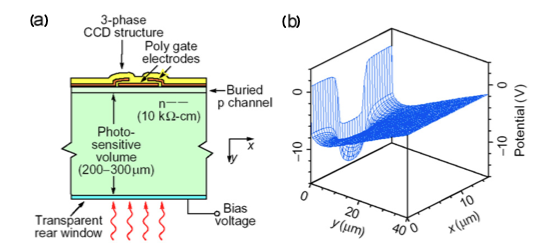}}
\caption{Cross section of a 250 $\mu$m thick CCD developed at Lawrence Berkeley National Laboratory, (a) layout of the three gates that form one pixel, (b) electrostatic potential (V) generated through the three gated phases is shown as function of depth (y axis) and one of the lateral directions (x axis). Figure from reference \cite{Oluseyo 2004}.} 
\label{fig:CCD layout}
\end{figure}

\begin{figure}
\resizebox{1\hsize}{!}{\includegraphics*{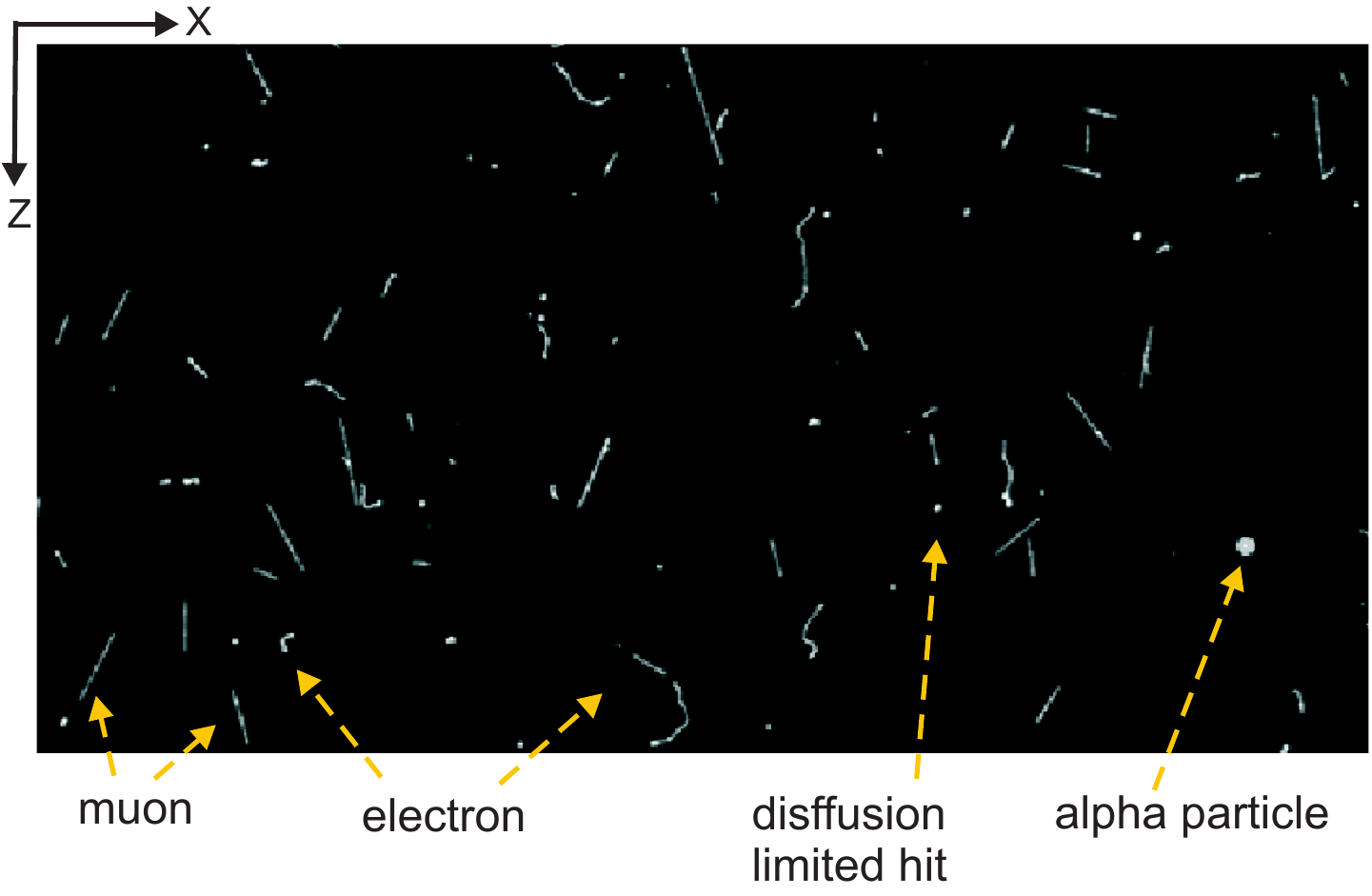}}
\caption{Compendium of images from recent measurement of background at sea level in a CCD (See text for an interpretation of the events).} 
\label{fig:compendium of background events}
\end{figure}

Figure \ref{fig:CCD layout} shows a scientific CCD developed by Lawrence Berkeley National Laboratory and characterized extensively at Fermilab for the DECam project \cite{Holland 2003, Estrada 2006}. Figure 1(a) shows the cross section of the layout of the three gates that compose one pixel. Figure 1(b) depicts the potential well generated under the gates in normal operation. Several million pixels CCD are fabricated on high resistivity silicon to maximize the depleted silicon volume and therefore  increase the near-IR photon response. CCDs with thickness of approximately $650\, \mu$m are available, and provide up to 5.2 grams of detector mass. The CCD is fully depleted with the use of a substrate voltage. The array is divided into square pixels of $15\,\mu {\rm m}$ by $15\,\mu {\rm m}$, which provides sufficient spatial resolution for efficient rejection of some of the background particles. An example of this characteristic is shown in Fig. \ref{fig:compendium of background events}, which presents a compendium of background events from measurements at sea level. 

Each particle produces a distinctive two-dimensional pattern in the CCD array. A muon is characterized by a straight-line track crossing the entire silicon volume. The small curved tracks are typical of energetic electrons produced by electromagnetic radiation. Alpha particles appear as big circular bright dots, due to the plasma effect they produce in the silicon \cite{Estrada 2011}. Finally, point events (energy deposited in a single pixel volume) are produced by the ionized charge that flows to neighbor pixels by diffusion. The coherent neutrino-nucleus scattering is expected to produce these kind of point events, as described in Section \ref{sec:identification of neutrino candidate events}.

\begin{figure}
\resizebox{1\hsize}{!}{\includegraphics*{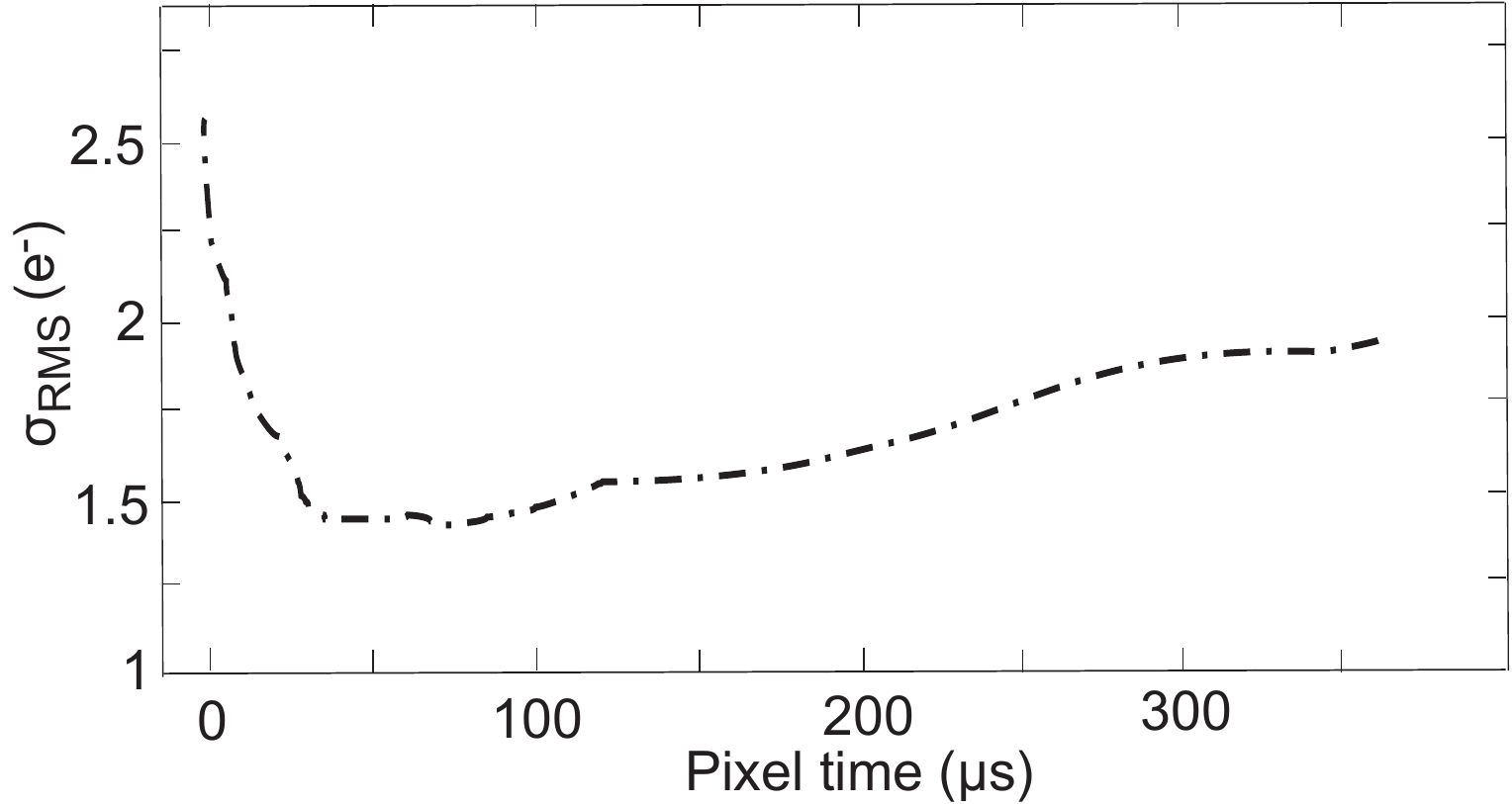}}
\caption{RMS pixel error ($\sigma_{RMS}$) caused by the output amplifier, as a function of pixel read-out time.} 
\label{fig:ccd noise}
\end{figure}

Besides the relatively large mass and high spatial resolution, these devices have a very small energy threshold, which is another attractive feature for neutrino detection. This characteristic is due to the small CCD readout noise, good charge transfer efficiency, and negligible dark-current contribution in a cooled system.
The readout noise is added to each pixel by the output amplifier during the charge packet readout. It has a Gaussian distribution with a standard deviation ($\sigma_{RMS}$) that depends on the readout time of the pixel, as shown in Fig. \ref{fig:ccd noise} (see \cite{Cancelo 2012,F. Moroni 2012} for a detailed analysis). Because of the interaction between $1/f$ and white noise, an optimum read-out noise with $\sigma_{RMS}=1.5\,{\text{e}}^{-}$ (equivalent to 5.5 eV of ionization energy) can be achieved using a pixel read-out time of $30 \,\mu {\text{s}}$. In what follows, it is assumed that this noise level is achieved during normal operation of the detector.

Current fabrication techniques and materials have yielded CCD detectors with dark-current generation below $2\,{\rm e}^{-}{\rm/day/pixel}$ when cooled at  123 K, and transfer inefficiencies below 15 ppm that  have negligible effect on the detection of low energy particles.

\section{NEUTRINO INTERACTION WITH MATTER}

A nuclear power reactor is a high flux source of electron antineutrinos ($\bar{\nu_e}$) with energies up to 12 MeV, approximately. At such energies, the largest probability for interaction with Si atoms is given by the coherent neutrino-nucleus neutral-current interaction. In this process, a neutrino of any flavor scatters off a Si nucleus transferring some energy in the form of a nuclear recoil. The SM cross section 
$\sigma$ for this process is \cite{Freedman 1974,Zuber 2011}
\begin{eqnarray}
 \frac{\text{d} \sigma}{\text{d} E_{\bar{\nu_e}} \text{d} E_{\text{rec}}}(E_{\bar{\nu_e}},E_{\text{rec}})=&&\frac{G_{F}^2}{8\pi}[Z(4\sin^2\theta_{W}-1)+N]^2\nonumber\\
&& \times
 M(2-\frac{E_{\text{ rec}}M}{{E_{\bar{\nu_e}}}^2})|f(q)|^2
\end{eqnarray}
where $M$, $N$ and $Z$ are, respectively, the mass, neutron number and atomic number of the nucleus, $E_{\bar{\nu_e}}$ and $E_{\text{rec}}$ are the incident neutrino and the nuclear recoil energy, $G_F$ is the Fermi coupling constant, $\theta_{W}$ is the weak mixing angle, and $f(q)$ is the nuclear form factor at momentum transfer $q$. According to \cite{Patton 2012}, $|f(q)| \approx 1$, within an uncertainty of a few percent. This is applicable for $E_{\bar{\nu_e}}<50$ MeV, where the momentum transfer ($q^2$) is small enough such that $q^2R^2<1$, where $R$ is the radius of the nucleus \cite{Scholberg 2006}. At small momentum transfers, the individual nucleon amplitudes are in phase and add coherently, so that the cross section increases by a factor of approximately $N^2$.

\begin{figure}
\resizebox{1\hsize}{!}{\includegraphics*{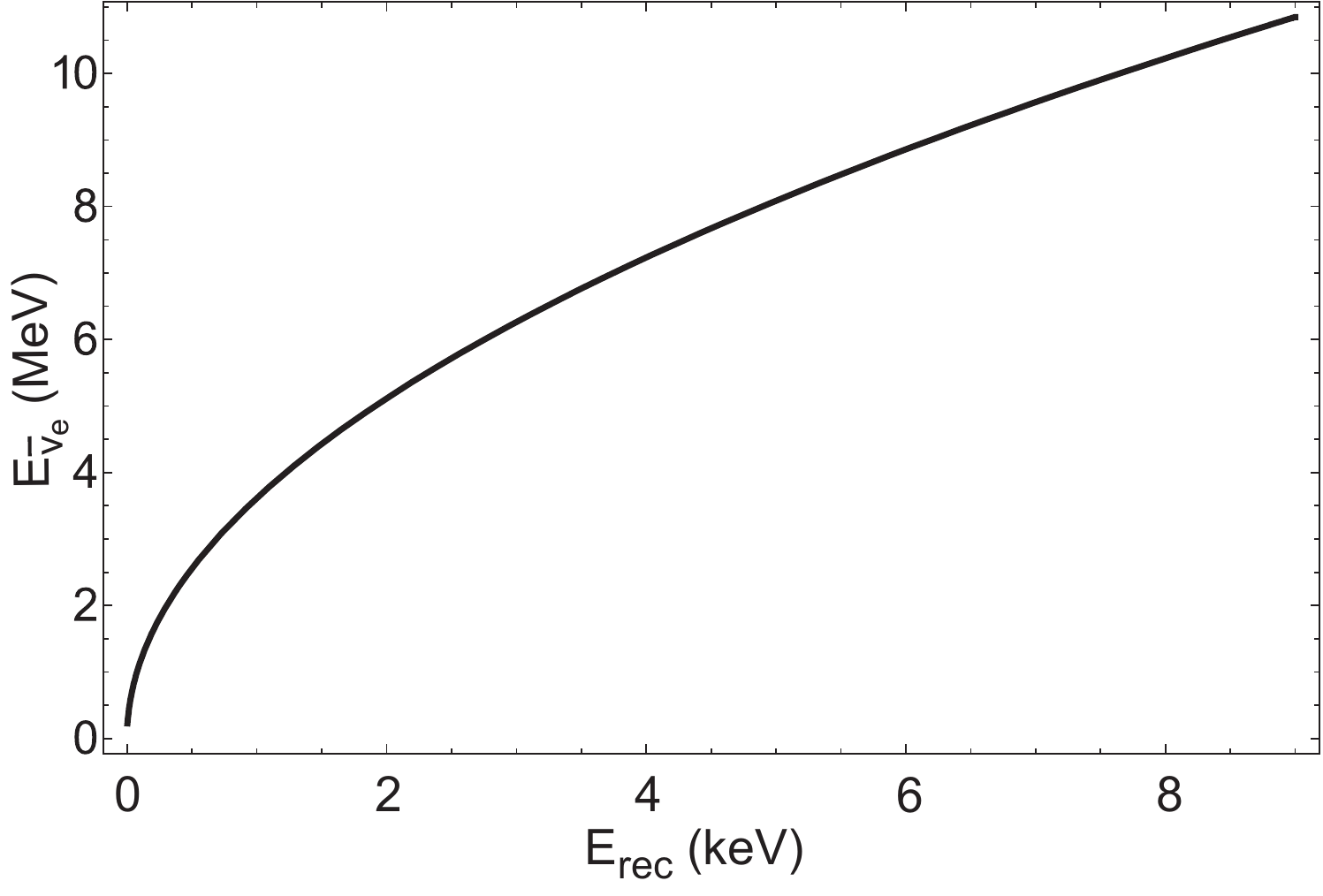}}
\caption{Neutrino energy as a function of maximum transferred energy to the Si nucleus.} 
\label{fig:maximum recoil energy}
\end{figure}

Although the cross section is enhanced by such coherence, elastic neutrino-nucleus scattering is difficult to observe because of the very small nuclear recoil energies. Figure \ref{fig:maximum recoil energy} depicts this relationship for silicon atoms, where the maximum event energy is max($E_{\text{ rec}}$)$=2E_{\bar{\nu_e}}^2/M$ (approximately 10 keV). Therefore, this kind of measurement requires  very sensitive detectors and a good characterization of the background.

\begin{figure}
\resizebox{1\hsize}{!}{\includegraphics*{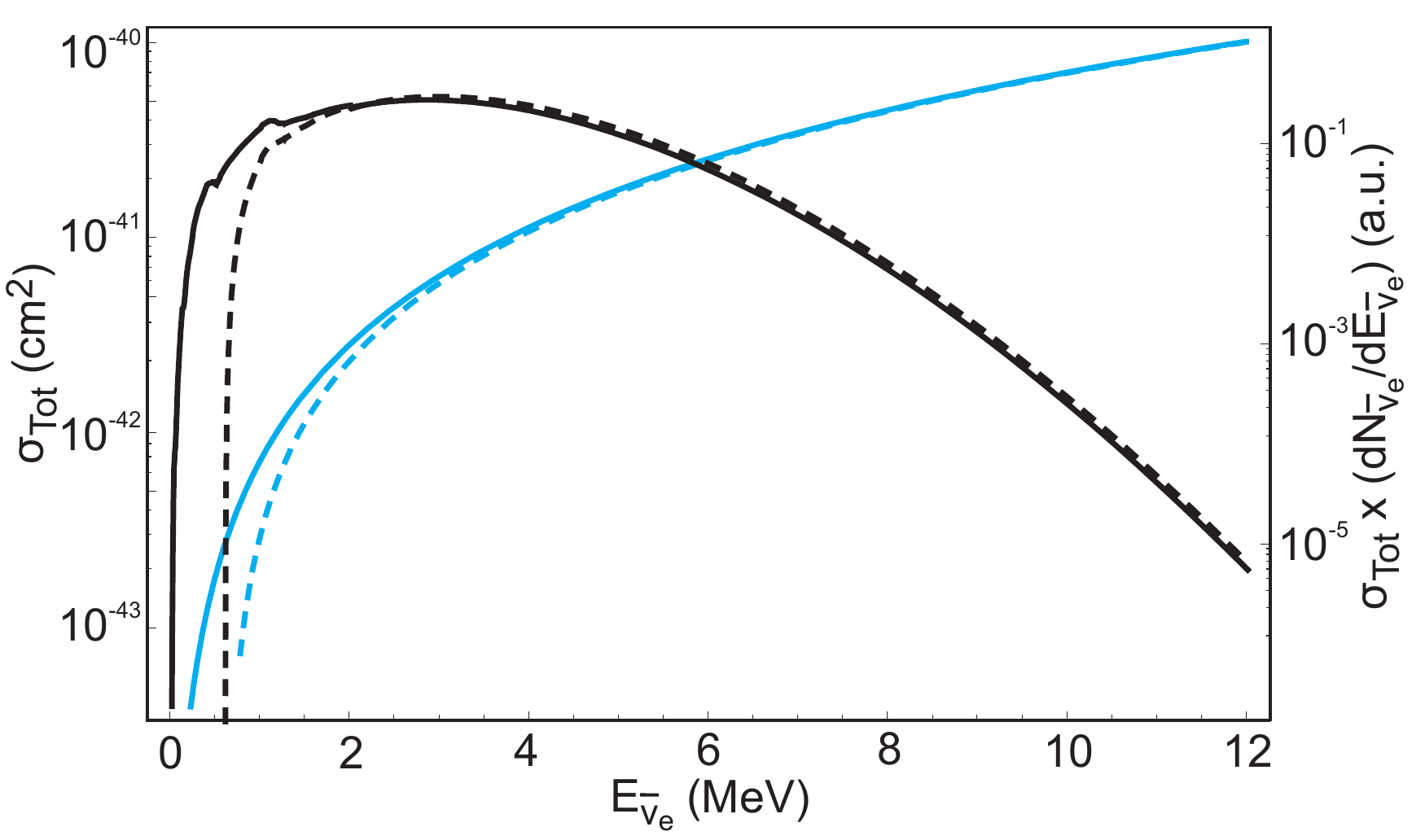}}
\caption{Total neutrino-nucleus coherent cross section $\sigma_{\rm T}$ for silicon from Eq. \ref{eq: total cross section for silicon} (light blue curve, left), and weighted by the reactor antineutrino spectrum (black curve, right). The dashed lines correspond to a threshold energy of $28$ eV, approximately $5 \sigma_{RMS}$ of the detector noise.} 
\label{fig:total cross section}
\end{figure}

The total cross section $\sigma_{\rm T}(E_{\bar{\nu_e}})$  for a mono-energetic neutrino source of energy $E_{\bar{\nu_e}}$ is given by 
\begin{equation}
	\sigma_{\rm T} (E_{\bar{\nu_e}}) =\frac{G_F^2}{4\pi}[Z(4\sin^2\theta_{W}-1)+N]^2E_{\bar{\nu_e}}^2 \nonumber
\end{equation}
that can be approximated by
\begin{equation}
\label{eq: total cross section for silicon} 
	\sigma_{\rm T} (E_{\bar{\nu_e}})  \approx 4.22\times 10^{-45}N^2 E_{\bar{\nu_e}}^2
\end{equation}
when $E_{\bar{\nu_e}}$ is expressed in MeV and $\sigma_{\rm T}$ in ${\rm cm}^2$. The total cross section $\sigma_{\rm T} $ for $^{28}$Si ($N=14$) is shown in light-blue trace in Fig. \ref{fig:total cross section} as a function of the neutrino energy $E_{\bar{\nu_e}}$, showing the small probability for interaction of low energy neutrinos with matter, and its strong dependence on incident energy. The total cross section $\sigma_{\rm T}$ weighted by the ${\bar{\nu_e}}$ energy spectrum from a reactor ($\text{d}N_{\bar{\nu_e}}/\text{d}E_{\bar{\nu_e}}$) is also depicted in Fig. \ref{fig:total cross section} using a black solid line, which is related to the probability of observing a reactor $\bar{\nu_e}$ of a given energy. The most probable event arises from neutrino energies between 2 MeV and 4 MeV. If the CCD threshold level is considered, the probability of detection is reduced, as depicted by the dashed curves in Fig. \ref{fig:total cross section}. In this case, the total cross section is calculated using a threshold of 28 eV, approximately 5 times the minimum RMS noise level $\sigma_{\rm RMS}$. These results, summarized in Fig. \ref{fig:total cross section}, suggest that the low threshold of the detectors is adequate for detecting $\bar{\nu_e}$ scattering.

\section{NEUTRINO SOURCE: NUCLEAR REACTOR}

Nuclear reactors emit about 3.1$\times 10^{16}$ $\bar{\nu_e}$/s per MW of thermal power, broadly distributed over energies up to 12 MeV, with a maximum between 0.5 MeV to 1 MeV. The antineutrinos come out isotropically from the core, so that the expected flux density at a distance $L$ is diminished by the factor $1/(4\pi L^2)$. 
At steady state operation, approximately 7.3 $\bar{\nu_e}$ ($N_{\bar{\nu_e}}$) are produced per reactor fission \cite{Wong 2007}. Many processes are involved in antineutrino production, but the  two major contributions are $\beta$ decays of fission fragments of the four fissile isotopes $^{235}$U, $^{238}$U, $^{239}$Pu, $^{241}$Pu  ($\approx$ 6.1 $\bar{\nu_e}$/fission), and neutron capture by $^{238}$U  ($\approx$ 1.2 $\bar{\nu_e}$/fission). The relative contribution from each source varies in different reactors, as well as in a single reactor during a burning cycle, resulting in antineutrino flux scenarios that differ by a few percent. Although such variations are clearly noticeable, they are small enough to provide an essentially model-independent analysis of any reactor neutrino experiment. In the following sections, each production mechanism is analyzed in more detail.

\subsection{Antineutrinos from fissile isotopes} 

\begin{figure}
\resizebox{1\hsize}{!}{\includegraphics*{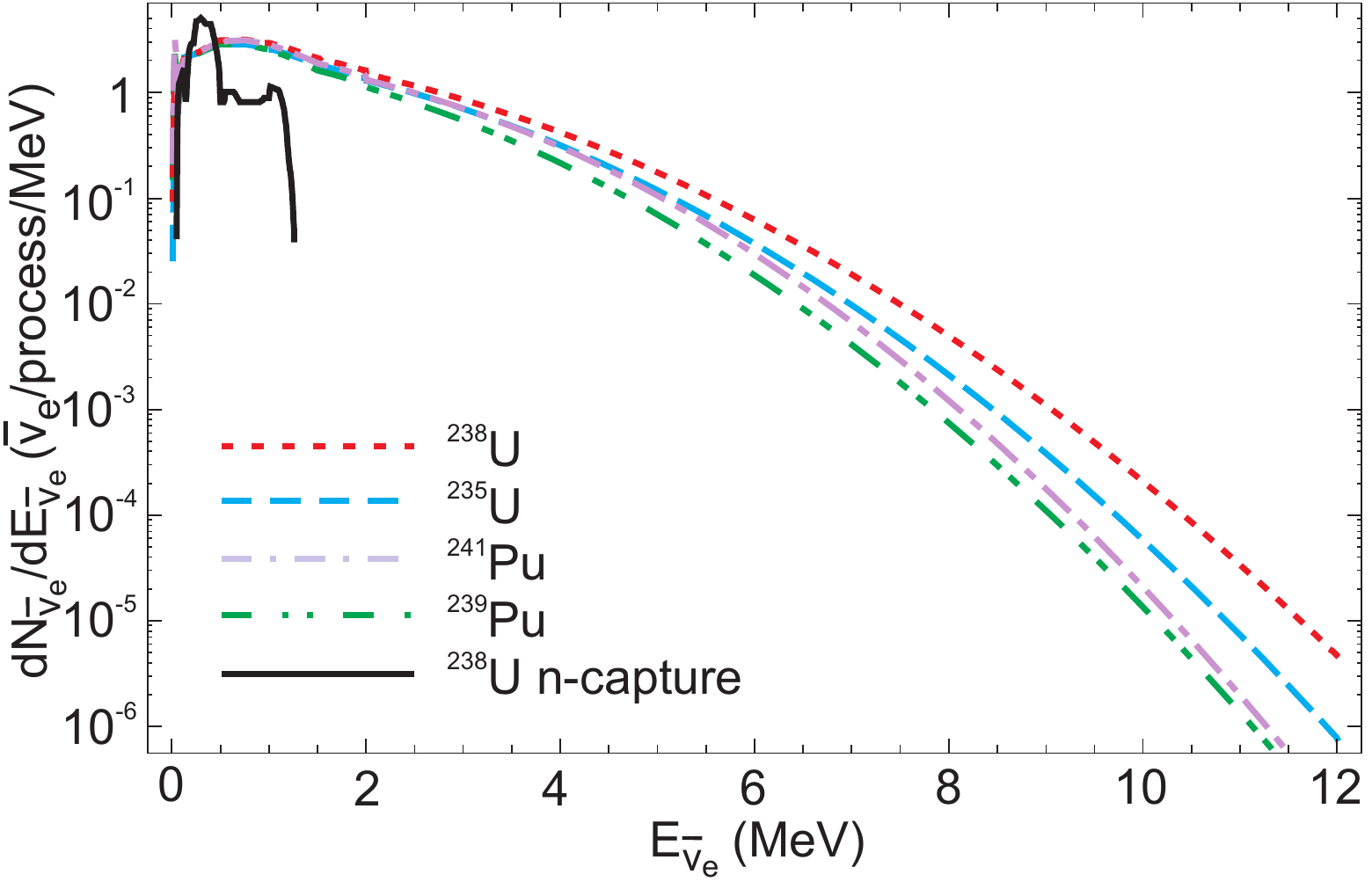}}
\caption{Antineutrino spectrum for each process.}    
\label{fig:antineutrino spectra}
\end{figure}

The $\bar{\nu_e}$ emitted in power reactors are predominantly produced through $\beta$-decays of the fission products, following the fission of the four dominant fissile isotopes: $^{235}$U, $^{238}$U, $^{239}$Pu, and $^{241}$Pu. Other fissile isotopes such as $^{236}$U, $^{240}$Pu, $^{242}$Pu, etc, contribute less than $0.1\,\%$ to the fissile isotope spectrum, and therefore can be neglected. Each isotope has a different $\bar{\nu_e}$ yield, $\bar{\nu_e}$ spectrum, and fission rate. Their content also changes during the fuel burning cycle, and leads to a small variation of the $\bar{\nu_e}$ flux and spectrum. This affects the total number $N_{\bar{\nu_e}}$ by a few percent, and can be ignored in a first order analysis of a CCD-based detector. 
The typical $\bar{\nu_e}$ yield per element fission, as well as their relative contributions per reactor fission are summarized in Table \ref{tab:neutrino contribution for each process}. The $\bar{\nu_e}$ spectrum produced through the fission of each isotope is depicted in Fig. \ref{fig:antineutrino spectra} in units of $\bar{\nu_e}$/MeV for each process. The antineutrino spectra are taken from \cite{Vogel 1981,Vogel 1989}. For energies above 2 MeV, the parametrization is represented by the model
\begin{equation}
\label{eq: neutrino spectrum} 
\text{d} N_{\bar{\nu_e}}/ \text{d} E_{\bar{\nu_e}} = \text{e}^{a_0 + a_1E_{\bar{\nu_e}} + a_2E_{\bar{\nu_e}}^2},
\end{equation}
where $a_0$, $a_1$, $a_2$ are the fitted parameters, with values shown in Table \ref{tab:spectra fitted constant}. For energies below 2 MeV, the antineutrino spectrum is given in tabulated form, and the values are listed in Table \ref{tab:low energy antineutrino spectra}.

\begin{table}[t]
\caption{Relative fission contribution and neutrino yield per fission for the four fissile isotopes and the $^{238}$U neutron capture. Typical values are given for integrated contributions.}
\label{tab:neutrino contribution for each process}  
\begin{ruledtabular}
\begin{tabular}{@{}llll}
Process  & Relative rate for  & Neutrino yield   & Neutrino yield\\
 &reactor fission &($N_{\bar{\nu_e}}$/process) &  ($N_{\bar{\nu_e}}$/fisision)\\
\colrule
$^{235}$U & $0.56$ & $6.14$ & $3.43$ \\
$^{238}$U & $0.08$ & $7.08$ & $0.56$ \\
$^{239}$Pu & $0.30$ & $5.58$ & $1.67$ \\
$^{241}$Pu & $0.06$ & $6.42$ & $0.38$ \\
$^{238}$U(n,$\gamma$) & $0.60$  & $2.00$ & $1.20$ \\
\end{tabular}
\end{ruledtabular}
\end{table}

\begin{table}[t]
\caption{Constants for Eq. (\ref{eq: neutrino spectrum}) for each fissile isotope.}
\label{tab:spectra fitted constant}      
\begin{ruledtabular}
\begin{tabular}{lccccc}
$a_i$  & $^{235}$U& $^{239}$Pu & $^{238}$U & $^{241}$Pu \\
\colrule
$a_0$ & \hphantom{$-$}1.260 & \hphantom{$-$}$ 1.0800$ & \hphantom{$-$}$ 1.500$ & \hphantom{$-$}$ 1.3200$ \\
$a_1$ & $-0.160$ & $-0.2390$ & $-0.162$ & $-0.0800$ \\
$a_2$ & $-0.091$ & $-0.0981$ & $-0.079$ & $-0.1085$ \\
\end{tabular}
\end{ruledtabular}
\end{table}

\begin{table}[t]
\caption{Low energy antineutrino spectra for each fissile isotope ($\bar{\nu_e}{\rm /MeV/fission}$ ).}
\label{tab:low energy antineutrino spectra}       
\begin{ruledtabular}
\begin{tabular}{lllll}
 E [MeV]  & $^{235}$U  & $^{239}$Pu & $^{238}$U & $^{241}$Pu  \\
\colrule
2 & 1.26 & 1.08 & 1.5 & 1.32\\
1.5 & 1.69 & 1.48  & 1.97  & 1.75\\
1 & 2.41 & 2.32 & 2.75 & 2.63\\
0.75 & 2.66 & 2.58 & 2.96 & 2.9 \\
0.5 & 2.66 & 2.63 & 2.91 & 2.82\\
0.25 & 2.16  & 2.08 & 2.18 & 2.14\\
0.125 & 1.98 & 1.99 & 2.02 & 1.85\\
6.25$\times{10^{-2}}$ & 0.61  & 0.64 & 0.65 & 0.59\\
3.12$\times{10^{-2}}$ & 0.35 & 2.13 & 1.32 & 3 \\
1.563$\times{10^{-2}}$ & 0.092 & 0.56 & 0.35 & 0.79\\
 7.813$\times{10^{-3}}$& 0.024  & 0.14 & 0.089 & 0.2\\
\end{tabular}
\end{ruledtabular}
\end{table}

\subsection{Antineutrinos from neutron capture in $^{238}$U}

The $^{238}$U content in power reactors nuclear fuel varies between $95\,\%$ to $97\,\%$. The $^{238}$U nuclei absorb approximately 0.6 neutrons per fission via the (n,$\gamma$) reaction: $^{238}$U + n $\Rightarrow$ $^{239}$U $\Rightarrow$ $^{239}$Np $\Rightarrow$ $^{239}$Pu. Two $\bar{\nu_e}$ are produced through $\beta$-decay of $^{239}$U.  This process contributes nearly $16\,\%$ to the total $\bar{\nu_e}$ flux. The $\bar{\nu_e}$ yield and rate per fission at the reactor are also summarized in Table \ref{tab:neutrino contribution for each process}. The energy of the antineutrinos produced by this process is below $1.3$ MeV, as shown in Fig. \ref{fig:antineutrino spectra} (black curve). A description of these processes can be found in \cite{Wong 2007}.

\begin{figure}
\resizebox{1\hsize}{!}{\includegraphics*{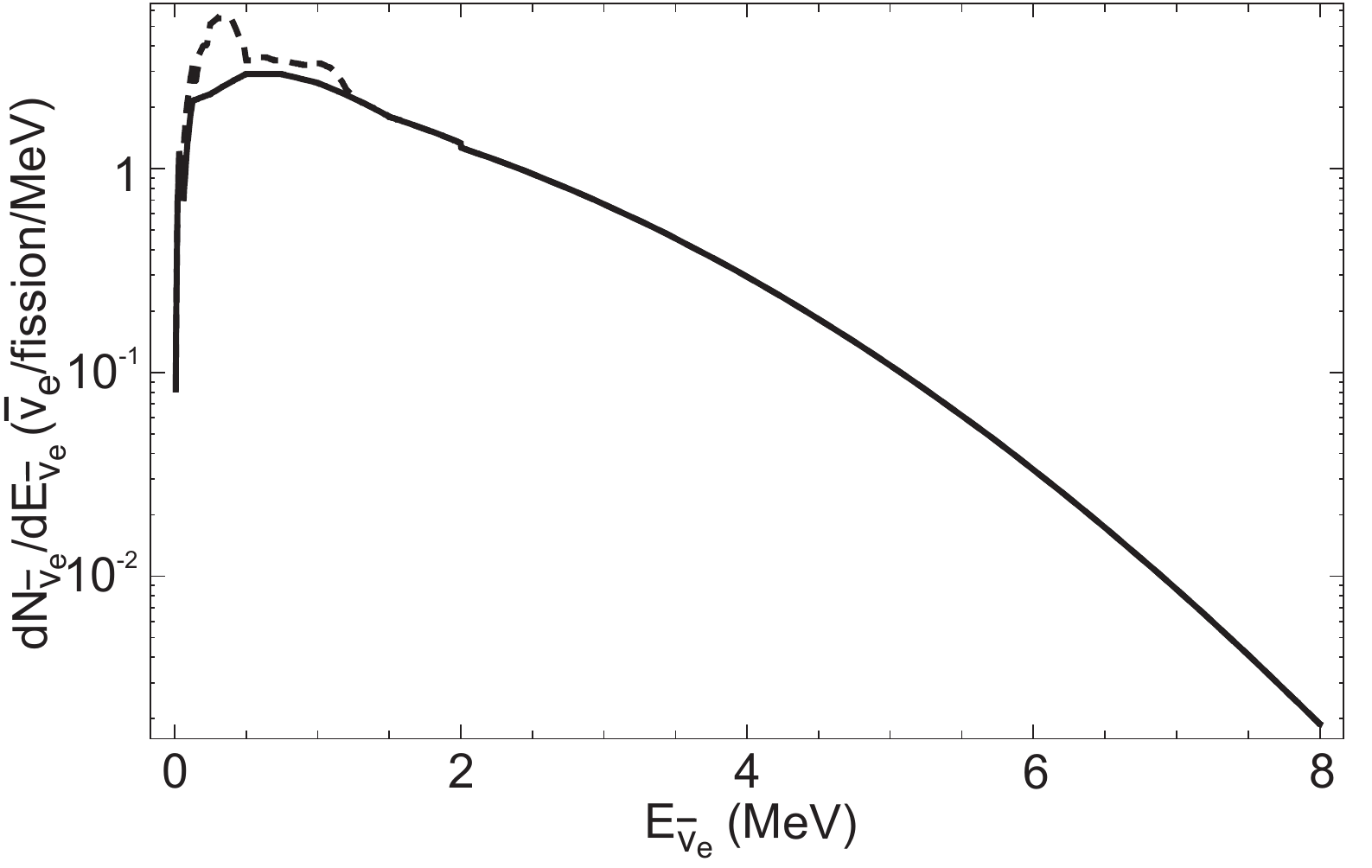}}
\caption{Total reactor $N_{\bar{\nu_e}}$ spectrum per fission in the reactor per MeV. The solid line reflects the fissile isotopes, and the dashed line, the sum of the fissile isotopes and the neutron capture by $^{238}$U.} 
\label{fig:total antineutrino spectrum}
\end{figure}

Figure \ref{fig:total antineutrino spectrum} depicts the total antineutrino spectrum per fission with (dashed line) and without (solid line) the contribution of the $^{238}$U capture mechanism. This graph can be translated to any experiment, by multiplying it by the number of fissions expected in the reactor, diminished by the distance factor.

\section{CCD EXPERIMENT AT REACTOR}

This work provides a preliminary analysis to forecast the expectations for the Coherent Neutrino-Nucleus Interaction Experiment  (CONNIE), currently under construction. The goal of CONNIE is the first unambigous detection of neutrino-nucleus coherent scattering using an array of CCD detectors in a radiation shield located 30 meters from the core of the Angra II reactor, which operates at a thermal power of 3.95 GW. This experiment is planned to be installed at the Almirante Alvaro Alberto Nuclear Central, in Angra Dos Reis, Brazil during 2014.

In steady-state operation, the neutrino flux produced by the reactor is $1.21\times 10^{20} \,\bar{\nu_e} /{\rm s}$ approximately, and the flux density at the detector ($L=30$ meters from the core) is $7.8 \times 10^{12}\,\bar{\nu_e} /{\rm cm}^{2}/{\rm s}$. These large numerical values justify the use of nuclear reactors as neutrino source for the CONNIE experiment.

The feasibility of neutrino detection close to a power nuclear reactor requires not only the estimation of the event rates and background noise, but also the proper identification of neutrino events. Once these parameters are known, the running time of the experiment to achieve a certain confidence level can be estimated.

\subsection{Event rate}

\begin{figure}[t]
\resizebox{1\hsize}{!}{\includegraphics*{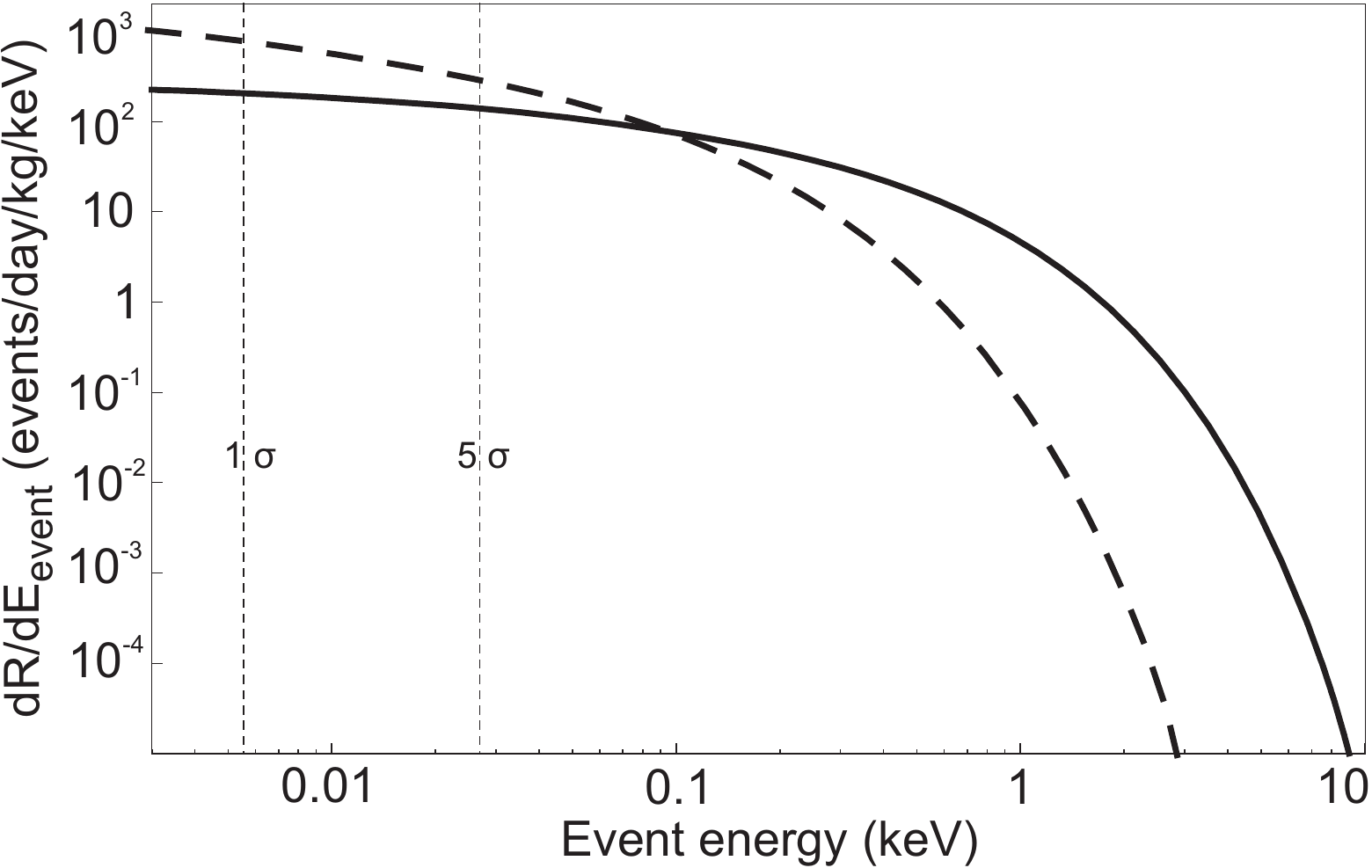}}
\caption{Energy spectra for events expected in silicon detectors: the nuclear-recoil energy spectrum (---); the spectrum for detectable events (-- --), using the quenching factor from Lindhand, \textit{et al.} \cite{Lindhard 1963,Chagani 2008}.} 
\label{fig:recoil energy spectrum}
\end{figure}

The product of the coherent scattering interaction is a nuclear recoil that ionizes electrons of Si atoms in the lattice, which are collected to form the event in the output image. Using the differential cross section, the total $\bar{\nu_e}$ spectrum and the $\bar{\nu_e}$ flux expected at the detector, the nuclear recoil spectrum $\text{d} R(E_{\text{rec}})/\text{d} E_{\text{rec}}$ is given by
\begin{eqnarray}
\frac{\text{d} R}{\text{d} E_{\text{rec}}}(E_{\text{rec}}) =&& N_t\int_{\sqrt{\frac{2E_{\text{rec}}^2}{M}}}^{\infty}\text{d} E_{\bar{\nu_e}}\frac{\text{d} N_{\bar{\nu_e}}}{\text{d} E_{\bar{\nu_e}}}(E_{\bar{\nu_e}}) \nonumber \\
&&\times  \frac{\text{d} \sigma}{\text{d} E_{\bar{\nu_e}} \text{d} E_{\text{ rec}}}(E_{\bar{\nu_e}},E_{\text{rec}}) 
 \label{eq: event spectrum} 
\end{eqnarray}
and the total rate for events $R$ in the energy range of the detector is given by
\begin{equation}
\label{eq: event rate} 
R = \int_{E_{th}}^{\infty}\text{d} E_{\rm rec}\frac{\text{d} R}{\text{d} E_{\text{ rec}}}(E_{\text{ rec}})\\
\end{equation}
where $\text{d} N_{\bar{\nu_e}}(E_{\bar{\nu_e}})/\text{d} E_{\bar{\nu_e}}$ represents the spectrum of neutrinos at the detector, $E_{th}$ is the detector's threshold energy, $N_t$ is the number of nuclei in the detector, and $\sqrt{2E_{\text{rec}}^2/M}$ is the minimum neutrino energy that can produce a recoil with energy $E_{\text{rec}}$. 

\begin{figure}[t]
\resizebox{1\hsize}{!}{\includegraphics*{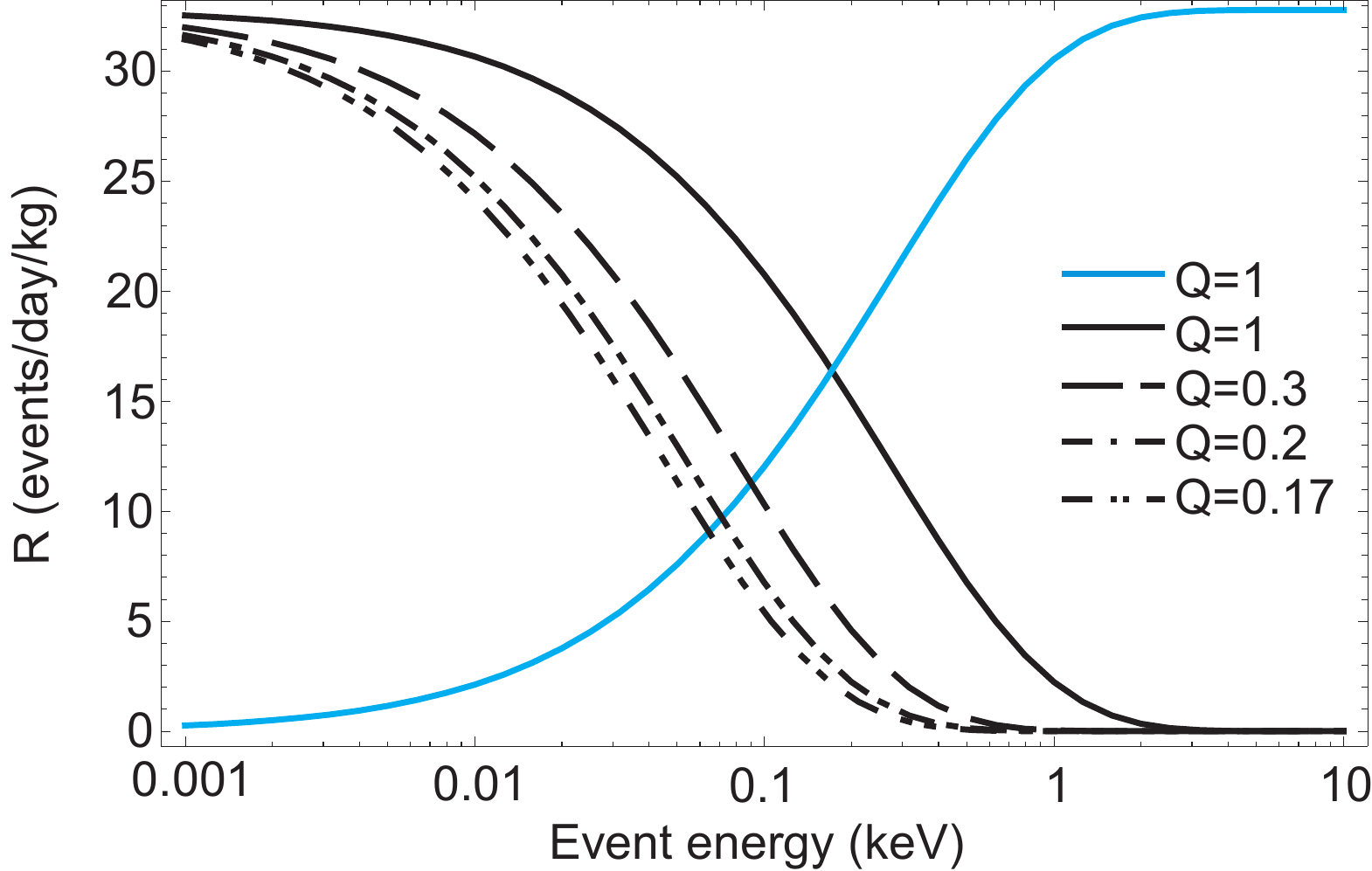}}
\caption{Total number of events as a function of the threshold energy for different quenching factors: $Q=1$, $Q=0.3$, $Q=0.2$ and $Q=0.17$ (black curves). The light-blue curve shows the total number of events as a function of the maximum detectable recoil energy using $Q=1$.} 
\label{fig:events rate above threshold}
\end{figure}

The results for $\text{d} R/\text{d} E_{\text{rec}}$ and $R$ from Eqs. (\ref{eq: event spectrum}) and (\ref{eq: event rate}) are shown in Fig. \ref{fig:recoil energy spectrum} and Fig. \ref{fig:events rate above threshold}, respectively. The nuclear recoil spectrum shown in Fig. \ref{fig:recoil energy spectrum} decreases rapidly with energy. Although events with $E_{\text{rec}}$ up to 10 keV are expected, any recoil for $E_{\text{rec}}> 3$ keV has a very low probability of occurrence. In fact, more than $96\,\%$ of the events occur for $E_{\text{rec}}< 2$ keV. This behavior can be also deduced from the integrated spectrum in Fig. \ref{fig:events rate above threshold} (light-blue curve), which represents the rate of events as a function of the upper limit in $E_{\text{rec}}$. Above 2 keV, the distribution becomes flat and there is essentially no significant increase in the event rate. This characteristic should be used to find the best energy cutoff to maximize the event to background ratio. The bounded energy range also provides some clues about the expected signature from $\bar{\nu_e}$-hits, as discussed in the next section. The use of heavier target materials result in an even shorter visible energy range.

\begin{figure}
\resizebox{1\hsize}{!}{\includegraphics*{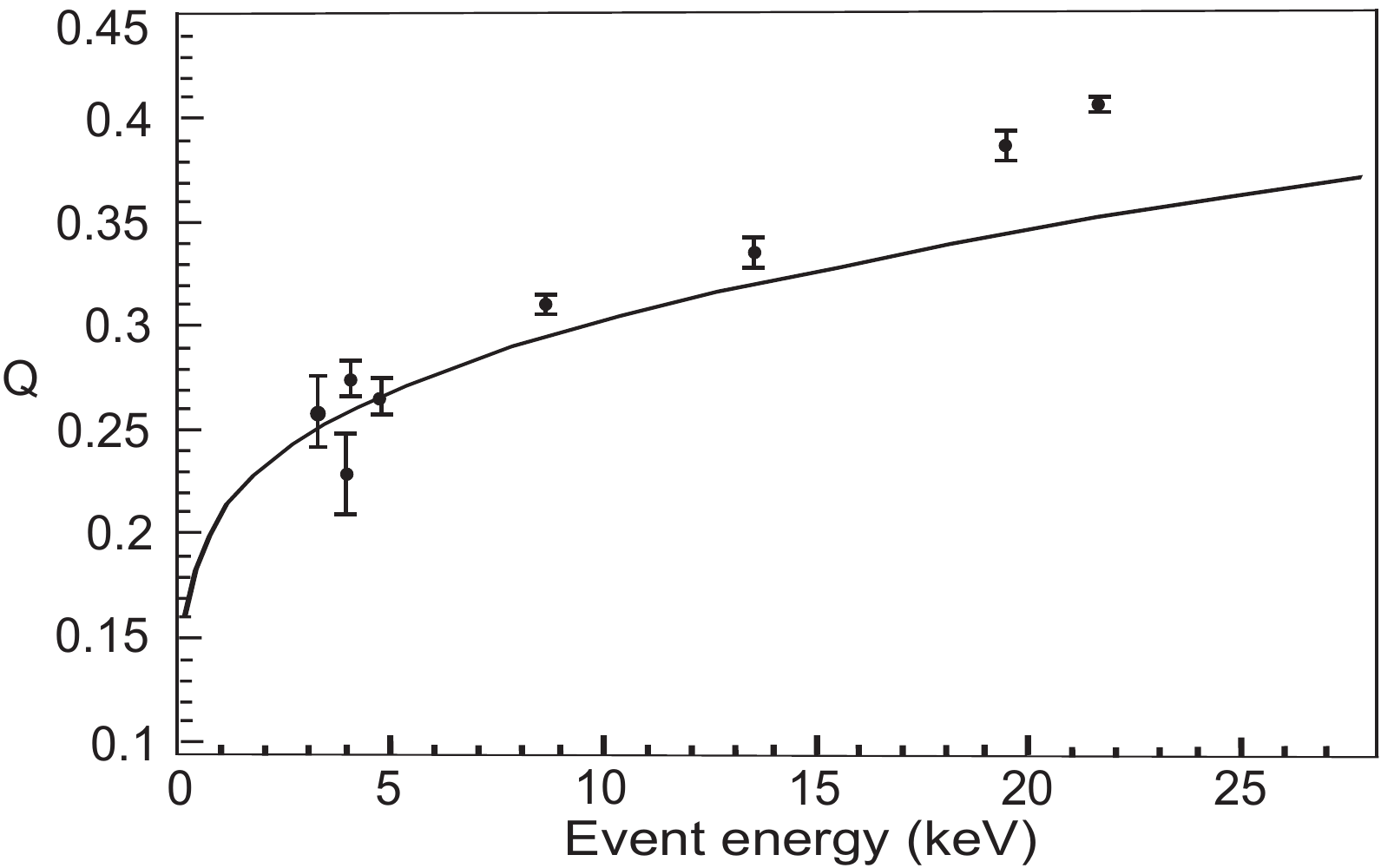}}
\caption{Silicon quenching factor. Measurements from \cite{Lewin 1996}, and theoretical prediction from \cite{Lindhard 1963}.} 
\label{fig:silicon quenching factor}
\end{figure}

Only a fraction of the nuclear recoil energy is converted into charge inside the Si detector, because part of the deposited energy results in phonons, contributing to the increase of the thermal energy of the system. The quantity that reflects the mean portion of the energy that is used in the ionizing process is the quenching factor $Q$. This factor has a strong dependence on energy and unfortunately it is not well known for energies $< 4$ keV, although there are several ongoing efforts to measure $Q$ in this energy range \cite{Tiffenberg 2013}. However, measurements for event energies $> 4$ keV \cite{Lewin 1996} agree with Lindhard's theory \cite{Lindhard 1963,Chagani 2008}. Figure \ref{fig:silicon quenching factor} shows the predicted silicon quenching factor by Lindhard, and the available measurements at different recoils energies. Taking into account the Lindhard $Q$ factor, the observable event energy spectrum is also shown as a dashed curve in Fig. \ref{fig:recoil energy spectrum}, indicating that the range of ionization energy is reduced to approximately 3 keV. The dependence on lower ionizing energies becomes stronger due to the reduction of $Q$ at small energy values. 

Figure \ref{fig:events rate above threshold} also depicts the total number of detected events for different threshold energies and different values of the quenching factor $Q$ (assuming $Q$ is a constant). Despite the quenching factor is not well known at low energies, the total number of events detected has a relatively weak dependence on it because of the very low noise of CCD devices. Table \ref{tab: expected number of events per kg} summarizes the number of events per day per kg of detector for different quenching factors and for two values of energy threshold $E_{th}$. The total number of events for zero energy threshold is expected to be 33 events per day per kg of silicon.

\begin{table}
\caption{Expected number of events for different quenching factors and threshold conditions, given in events/kg/day.}
\label{tab: expected number of events per kg} 
\begin{ruledtabular}
\begin{tabular}{@{}lllll}
 $E_{th}$& $Q=1$  & $Q=0.3$ & $Q=0.2$ & $Q=0.17$  \\
\colrule
1$\sigma_{RMS} (5.5eV)$ & 30 & 29 & 28 & 27\\
5$\sigma_{RMS} (28eV)$ & 28 & 22 & 18 & 17\\
\end{tabular}
\end{ruledtabular}
\end{table}

\subsection{Identification of neutrino candidate events}
\label{Sec:NeutrinoCandidates}
\label{sec:identification of neutrino candidate events}

The low energy nuclear recoil signature in the CCD corresponds to a diffusion limited hit, which means that the observed charge is generated in a volume smaller than the pixel size, and the event is formed only by the diffusion of the free charge in the silicon \cite{Barreto 2012}. 

When the charge is free to move in the Si lattice, the diffusion and drift mechanisms define its final lateral dispersion before it is trapped by the electric potential well under the gates. As it was explained in section \ref{sec: high resitivity CCDs}, the lateral barriers extend approximately 10 $\mu$m in depth ($y$ axes). Beyond this point, the electric field in the entire silicon bulk is uniform along the $x$ and $z$ axis, and varies only as a linear function of $y$ (a detailed electrostatic description of the devices can be found in \cite{Holland 2003}). The net result is that most of the carriers reach the well of the gate in the same pixel in which they were generated, and only a small fraction transverse to adjacent pixels. 

\begin{figure}
\resizebox{1\hsize}{!}{\includegraphics*{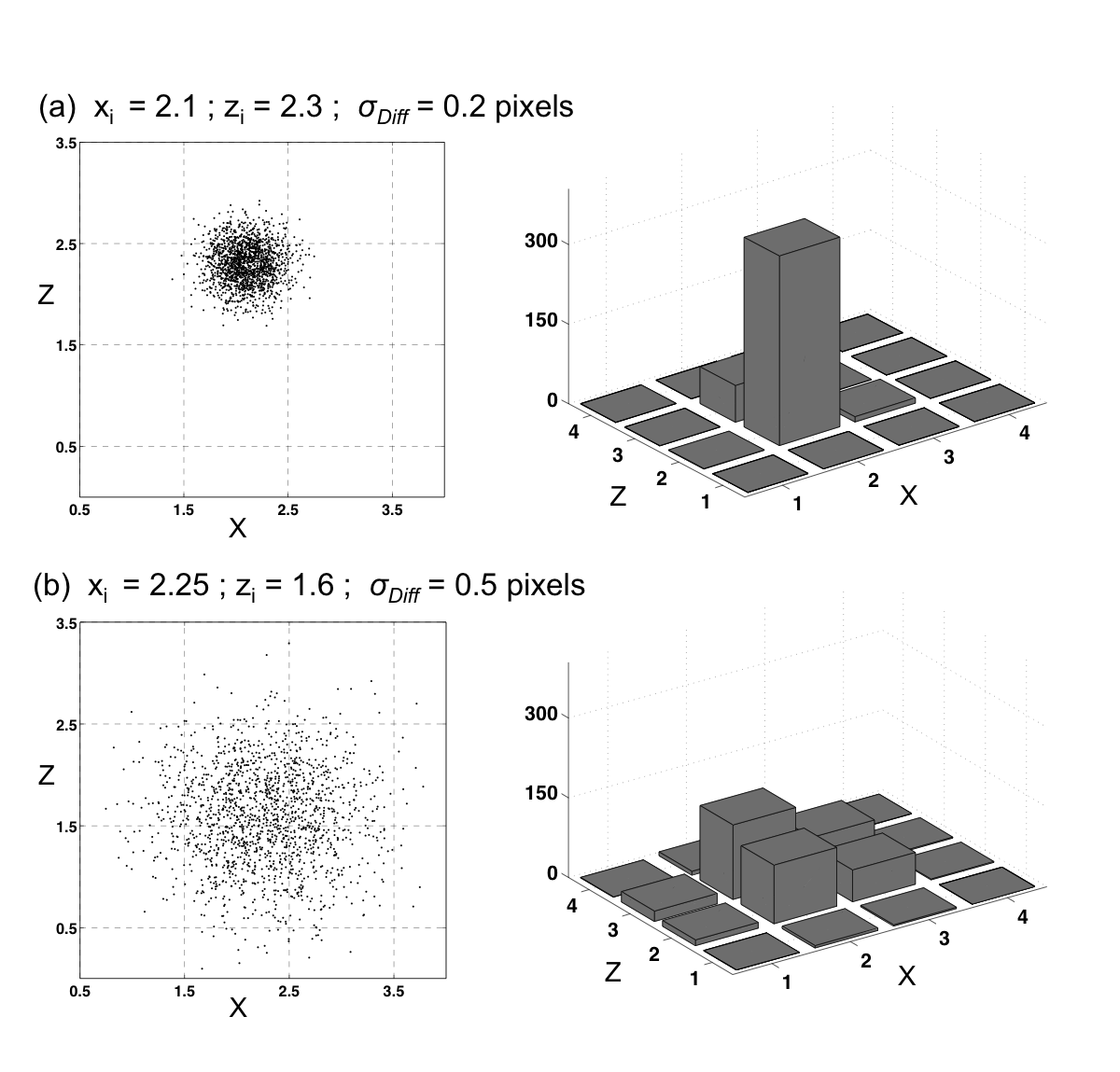}}
\caption{Two simulated neutrino events generated at different depths of the detector and at different relative position in the pixel. (a) $\bar{\nu_e}$-event interacting close to the gates of the detector, with $\sigma_{\rm Diff}=0.2$ pixels, and (b) $\bar{\nu_e}$-event interacting close to the back (large $y$) of the detector, where $\sigma_{\rm Diff}=0.5$ pixels. The $x_i$ and $z_i$ values are the coordinates of the point of origin of the events in the array. } 
\label{fig:simulated events}
\end{figure}

The ``pixelation'' of the detector plays an important role in the final shape of the expected event, giving a 2D stepped representation of the Gaussian distribution expected from diffusion. Due to the small number of pixels that form the event, the shape of the stepped distribution depends strongly on the initial lateral position of the charge relative to the boundaries of the pixel. Figure \ref{fig:simulated events} shows the effect of diffusion and pixelation on a simulated neutrino event produced at different depths and lateral positions. The energy of the event is 1.6 keV and it is simulated as interacting very close to the gates of the CCD in the $y$ axes (standard deviation of the diffusion distribution: $\sigma_{\rm Diff} = 0.2$ pixels) with a lateral position of $(x_i,z_i)$=(2.1,2.3) pixels in the array, in Fig. \ref{fig:simulated events}(a), and at $y\approx 250\,\mu$m at the back of the detector ($\sigma_{\rm Diff} = 0.5$ pixels) with $(x_i,z_i)$=(2.25,1.6) in Fig. \ref{fig:simulated events}(b). A detailed description of the shape of diffusion limited hits can be found in \cite{Barreto 2012,Tiffenberg 2013,Estrada 2011,F. Moroni 2013}.

The energy calibration of CCDs can be performed using several standard procedures. The most intuitive technique is using an X-ray source, specially $^{55}$Fe. A full description of the  procedure can be found in \cite{Janesick 2001}, and the calibration for these detectors in \cite{Barreto 2012,Tiffenberg 2013}.

\subsection{Running conditions and forecast}

The current version of the CONNIE detector is based on 10 CCD running in parallel. The CCD setup has capacity to read CCD of any thickness, and array sizes of up to approximately 6 cm by 6 cm. The system was designed for easy on-site replacement of detectors. After a preliminary operating stage, ten 5.2 g CCD units are planned to be running, summing 52 grams of detecting mass. The final spectrum and rate of events can be calculated from Figs. \ref{fig:recoil energy spectrum} and \ref{fig:events rate above threshold},  respectively. The number of expected events is 1.716 per day, totaling 626 events per year. Different conditions for anticipated $Q$ and threshold energies are summarized in Table \ref{tab: expected number of events in CCD setup}.

\begin{table}[t]
\caption{Expected number of events in the CCD array, for a mass of 52 g, for different quenching factors and threshold conditions. Results are given in units of events/day (events/year).}
\label{tab: expected number of events in CCD setup}       
\begin{ruledtabular}
\begin{tabular}{@{}lllll}
 $E_{th}$ & $Q=1$  & $Q=0.3$ & $Q=0.2$ & $Q=0.17$\\
\colrule
1$\sigma_{RMS}$ (5.5eV) & 1.56 (569) & 1.5 (547) & 1.46 (532) & 1.4 (511)\\
5$\sigma_{RMS}$ (28eV) & 1.46 (533) & 1.14 (416) & 0.94 (343) & 0.9 (328) \\
\end{tabular}
\end{ruledtabular}
\end{table}

To provide a first-order calculation of the expected running time, the result can be viewed as a counting experiment for a signal, expected to be higher than the Poisson fluctuation of the background at some given confidence level, for any specified range of energy.

Available bibliography shows that the count rate from background events in the low energy region at sea level using passive shield can be reduced to $\approx$ 600 events/KeV/day/Kg \cite{Heusser 1995}, assuming that the material of the shield has a low level of radiative contamination. Similar rates of background have been reached using similar configurations of CCD at shallow depth (30 m.w.e) in the Minos tunnel at Fermilab, and deep underground (600 m.w.e) at Snolab \cite{Barreto 2012,Tiffenberg 2013}.

Figure \ref{fig:events rate above threshold} shows that for $Q < 0.3$ almost all event have an ionization energy of $<300$ eV. The energy range of our interest lies between 28 eV ($5 \sigma_{RMS}$) and 300 eV.  Assuming $Q = 0.2$, the event rate from Table \ref{tab: expected number of events in CCD setup} yields 0.94 events/day for a 52 g array of CCD.

The background noise can also be scaled by the mass of the detector and by the energy interval resulting in a rate of 8.5 events/day. A signal-to-noise ratio defined as $0.94 T/2.91 \sqrt{T} = 0.32\sqrt{T}$ where $T$ is the running time in days, can be used to obtain the corresponding confidence value.
Therefore, the number of days running the experiment to achieve a certain confidence level (CL) can be computed, and some values for several CL values are listed in Table \ref{tab: running time}.

\begin{table}[t]
\caption{Expected running time for achieving different CL [PDG].}
\label{tab: running time}       
\begin{ruledtabular}
\begin{tabular}{@{}ll}
CL [\%] & $T$ [days]\\
\colrule
80.00 & 07 \\
90.00 & 16 \\
95.00 & 27 \\
98.00 & 44 \\
99.87 & 87 \\
\end{tabular}
\end{ruledtabular}
\end{table}

\section{CONCLUSION}

The capabilities of Charge Coupled Devices to detect coherent neutrino-nucleus scattering interaction has been demonstrated. The small threshold achieved on these devices allows the detection of small depositions of energy, in particular, nuclear recoils from neutrino scattering. On this energy range, the interaction occurrence is enhanced by coherence and therefore the neutrino signal can be observed using a system with moderate detecting mass. 

The basis for a coherent neutrino nucleus scattering experiment at a nuclear reactor have been also reviewed. The article shows that a neutrino signal of 626 events per year can expected in a CCD array of 52 g, with a certainty greater than 99$\%$ over the background fluctuation after ninety days of measurements.

\section{Acknowledgments}
The authors wish to thank Dr. Tom Ferbel, University of Rochester, for his review of an early draft, and his many suggestions.

\end{document}